\documentclass[9pt,conference]{IEEEtran}
\IEEEoverridecommandlockouts

\usepackage{cite}
\usepackage{amsmath,amssymb,amsfonts}
\usepackage{algorithmic}
\usepackage{graphicx}
\usepackage{textcomp}
\usepackage{xcolor}
\usepackage{url}
\def\BibTeX{{\rm B\kern-.05em{\sc i\kern-.025em b}\kern-.08em
    T\kern-.1667em\lower.7ex\hbox{E}\kern-.125emX}}

\usepackage{booktabs}
\usepackage{pifont}
\usepackage{multirow}
\usepackage{wrapfig}
\usepackage{diagbox}
\usepackage{amsmath}
\usepackage{threeparttable}

\newcommand{\tabincell}[2]{\begin{tabular}{@{}#1@{}}#2\end{tabular}}

\begin{document}

\title{Controllable Text-to-Speech Synthesis with Masked-Autoencoded Style-Rich Representation\\}

\author{
\IEEEauthorblockN{Yongqi Wang$^{1\star}$ , Chunlei Zhang$^{2\star}$, Hangting Chen$^{2}$, Zhou Zhao$^{1}$, Dong Yu$^{2}$}
\IEEEauthorblockA{$^1$ Zhejiang University $^2$ Tencent AI Lab}
\thanks{${\star}$ Equal contribution}.
}


\maketitle

\begin{abstract}
Controllable TTS models with natural language prompts often lack the ability for fine-grained control and face a scarcity of high-quality data. We propose a two-stage style-controllable TTS system with language models, utilizing a quantized masked-autoencoded style-rich representation as an intermediary. In the first stage, an autoregressive transformer is used for the conditional generation of these style-rich tokens from text and control signals. The second stage generates codec tokens from both text and sampled style-rich tokens. Experiments show that training the first-stage model on extensive datasets enhances the content robustness of the two-stage model as well as control capabilities over multiple attributes. By selectively combining discrete labels and speaker embeddings, we explore fully controlling the speaker’s timbre and other stylistic information, and adjusting attributes like emotion for a specified speaker. Audio samples are available at \url{https://style-ar-tts.github.io}.
\end{abstract}


\section{Introduction}

Controllable text-to-speech (TTS) systems aim to generate high-fidelity speech while allowing control over various style attributes of the synthesized speech, such as speaker timbre, pitch level and variation, emotion, acoustic environment, etc. Due to its promising applications in digital media production and human-computer interaction, controllable TTS has been attracting growing interest in the machine learning community with a substantial amount of research working on it \cite{guo2023prompttts, leng2023prompttts, ji2024textrolspeech,yang2024instructtts,zhou2024voxinstruct}.

Despite the extensive research on this topic, controllable TTS still faces some unsolved challenges: \textbf{(1) Control Interface Issue}. Most existing works use natural language prompts as a medium of style control, which is friendly for non-professional users. However, style descriptions with natural language tend to be broad and coarse-grained, making it difficult to precisely control specific attributes. Moreover, the rich diversity of natural language brings more challenges to modeling the relationship between style attributes and prompts. It is also difficult to fully encompass the user instructions in real-world scenarios, restricting the application of these methods. \textbf{(2) Data Issue}. The training of well-performed TTS systems relies on high-quality speech corpora, which are often limited in both data volume and stylistic diversity. When using natural language as the control interface, the additional cost of generating prompt sentences further restricts the data size. Present controllable TTS datasets like \cite{guo2023prompttts, ji2024textrolspeech} are often limited to hundreds of hours. This constraint puts challenges on learning precise control abilities and improving generation diversity.

In this paper, we propose a fine-grained controllable TTS system. In contrast to natural language prompts, We divide the value ranges of various stylistic attributes of speech into multiple intervals, each represented by a label, and use these labels as conditional inputs to achieve fine-grained control. By selectively combining these labels with speaker embeddings, we can generate new speaker timbre while controlling other attributes, or adjust certain attributes such as emotion for a given speaker.

Our controllable TTS system adopts a two-stage generation paradigm using two language models (LMs), with a style-rich representation as an intermediate output. We adopt a masked autoencoder (MAE) which learns to capture diverse style information by reconstructing mel filterbank from the encoded content input and masked fbank. The features extracted by the style encoder of the trained MAE are then discretized and used as an intermediary of the TTS pipeline. Each of the two stages relies on a decoder-only transformer. The first stage generates style-rich tokens conditioned on content and control signals, while the second stage generates codec tokens from the content input and the predicted style-rich tokens. Due to low dependence on high-quality corpora, the style-rich token generation phase can scale up to a large amount of data, boosting control capability and generation diversity; while in the codec token generation stage, a relatively small amount of data is sufficient to learn how to reconstruct codec tokens from content and style units, addressing the issue of high-quality data scarcity. To enhance the control accuracy of fine-grained attributes, we investigate classifier-free guidance in the style-rich token generation stage. Experiments indicate that our model achieves good style control ability while keeping decent audio quality and content accuracy.

\begin{figure*}[t]
\begin{center}
\includegraphics[width=0.7\textwidth]{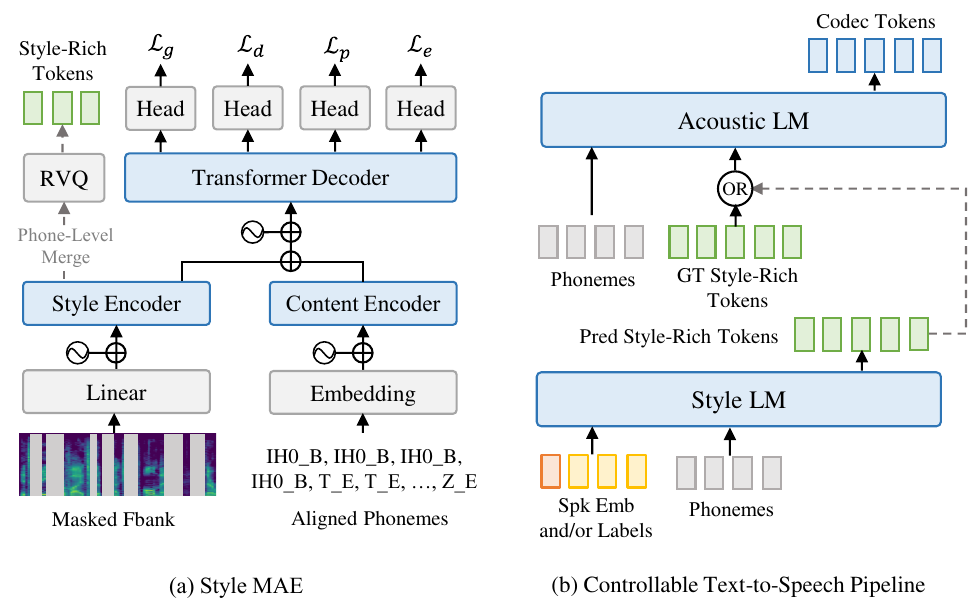}
\end{center}
\caption{Model overview of our controllable TTS system. Figure (a) shows the architecture of the style MAE. Figure (b) illustrates the two-stage controllable TTS pipeline. The gray dashed lines represent paths that occur only during inference.}
\label{fig:main}
\end{figure*}

\section{Method}

\subsection{Overview}

Our controllable TTS system consists of two major stages with a discrete style-rich token as an intermediate representation. This style-rich representation is from a transformer-based MAE as illustrated in figure~\ref{fig:main} (a), which learns to capture style information including speaker timbre, prosody, and acoustic environment in the speech with a mask-reconstruction paradigm. The style-rich tokens of a speech clip can be extracted with the style encoder of the pre-trained MAE followed by a residual vector quantizer (RVQ) trained individually. The two stages of TTS are (1) \textbf{style-rich token (ST) generation}, which generates style-rich tokens conditioned on content phonemes and style controlling signals including discrete labels and / or continuous speaker embeddings; and (2) \textbf{codec token (CT) generation}, which generates codec tokens conditioned on content phonemes and style-rich tokens, where the style-rich tokens are either extracted from ground truth speech or predicted by the former stage. The generated codec tokens are then used to reconstruct the waveform with the codec decoder. Each of the two stages relies on a decoder-only transformer to conduct LM-style generation, as illustrated in figure~\ref{fig:main} (b). We provide details of these modules respectively in the following subsections.

\subsection{Style masked autoencoder and feature tokenization}

The style masked autoencoder aims to learn to extract style information like speaker timbre, prosody, and acoustic environment by reconstructing mel filterbanks from masked ones and an additional content input with reconstruction and several auxiliary losses. Its architecture is illustrated in figure~\ref{fig:main} (a). The two branches of input, which are masked fbanks and a temporal-aligned phoneme sequence where each phoneme is duplicated by its duration, are processed by two encoders separately. Both the style encoder and content encoder are multi-layer transformer encoders. The output of the two encoders together with sinusoidal positional embedding are added and fed to the transformer decoder. 

Following \cite{huang2023prosody}, we append four different linear heads at the end of the decoder for output projection used for different optimization objectives. The four objectives are (1) reconstruction loss $\mathcal{L}_r$: mean square loss between the masked fbank patches and the output of the reconstruction head; (2) contrastive loss $\mathcal{L}_c$: InfoNCE loss to maximize the similarity between the head output and the corresponding fbank patch, while minimizing its similarity with non-corresponding fbank patches; (3) pitch classification loss $\mathcal{L}_p$ and (4) energy classification loss $\mathcal{L}_e$, which are cross-entropy losses calculated on log-scale fundamental frequency (f0) and the L2-norm of the amplitude spectrogram from short-time Fourier transform, respectively, both of which are frame-level and binned to 256 scales. The final loss is a linear combination of the four losses:
\begin{equation}
    \mathcal{L} = \lambda_r \mathcal{L}_r + \lambda_c \mathcal{L}_c + \lambda_p \mathcal{L}_p + \lambda_e \mathcal{L}_e
\end{equation}

where $\lambda_r$ = 10, and $\lambda_c$, $\lambda_p$, $\lambda_e$ are all 1. Intuitively, this design enables the MAE to extract content information from the encoded feature of the aligned phonemes, while extracting style information from the encoded feature of the masked fbank for reconstruction. Once the MAE finishes training, its style encoder should be able to capture various style information from speech.

To reduce the sequence length for language modeling and eliminate redundant information in the style features, we conduct phone-level merge by averaging the frame-level features in the range of each phoneme. After that, we train an RVQ with 3 codebooks independently over the phone-level style features for discretizing the style-rich representation for LM-style modeling. Note that such an architecture and training approach cannot fully prevent content information from leaking into the representations extracted by the style encoder, as it does not include a suitable bottleneck or supervisory signal to achieve this. This is why we refer to it as \textit{style-rich} token rather than \textit{style} token, and include content phonemes as part of the input in the ST generation stage. Nevertheless, this does not hinder the effectiveness of this representation in subsequent TTS applications.

\subsection{Two-stage LM-style controllable text-to-speech}

We use a decoder-only transformer for autoregressive generation for each of the two stages. Specifically, we adopt the multi-scale transformer in \cite{yanguniaudio} as the backbone model, which utilizes a stacked global-local transformer architecture to handle multi-codebook token modeling and has exhibited remarkable capabilities in audio synthesis. During training, the conditional inputs and target outputs are concatenated into a single sequence and fed to the transformer, with each part having a modality-specific \textit{start} and \textit{end} token at both ends. The LMs model the conditional distribution using next-token prediction with cross-entropy loss calculated on the target output part. 

\textbf{ST Generation} In the first stage, we adopt a style LM to generate style-rich tokens from phonemes and control signals. This procedure can be formulated as:
\begin{equation}
    {\mathrm{P}}(\mathbf{s}) = \prod_{t=1}^T \prod_{i=1}^N {\mathrm{P}}(s_t^i|\tau, c, \mathbf{s}_{<t}, \mathbf{s}_{t}^{<i}; \theta_s)
\end{equation}
where $\mathbf{s}$, $\tau$, $c$, and $\theta_s$ are style-rich tokens, phonemes, control signals, and model parameters, respectively. Here, the control signals can be a speaker embedding and / or discrete control labels. For discrete control labels, we include \textit{age}, \textit{gender}, \textit{pitch mean} for average pitch, \textit{pitch std} for the extent of pitch variation, emotion represented by \textit{arousal}, \textit{dominance}, and \textit{valence}, \textit{SNR} for signal-noise rate, and \textit{C50} for reverberation level. These labels are denoted by extracting attribute values with some tools and binning them to different levels. We can use all these labels to generate speech with a new speaker, or combine part of them like emotion labels with a speaker embedding to adjust these attributes on the basis of a reference speaker. The training data of this stage can be scaled up to large corpora to achieve higher style diversity and control accuracy.

\textbf{CT Generation} In the second stage, we adopt an acoustic LM to generate codec tokens from phonemes and style-rich tokens. No additional control signal is used in this stage, as we assume that the style information is carried by the style-rich tokens. During training, the model takes ground truth style-rich tokens and learns to reconstruct codec tokens of the speech. In inference, the style-rich tokens can be either ground truth ones for speech reconstruction, or predicted ones from the former stage for controllable TTS. This procedure can be formulated as:
\begin{equation}
    {\mathrm{P}}(\mathbf{a}) = \prod_{t=1}^T \prod_{i=1}^N {\mathrm{P}}(a_t^i|\tau, \mathbf{s}, \mathbf{a}_{<t}, \mathbf{a}_{t}^{<i}; \theta_a) .
\end{equation}
where $\mathbf{a}$, $\tau$, $\mathbf{s}$, and $\theta_a$ are codec tokens, phonemes, style-rich tokens, and model parameters, respectively. We observe in our experiment that several hundred hours of data are sufficient for the model to learn to reconstruct speech of decent quality from phoneme and style-rich tokens, therefore addressing the scarcity issue of high-quality corpora for controllable TTS.

\subsection{Classifier-free guidance}

We observe that for attributes with distinct differences among categories (like gender), simply adding the label to the prefix condition sequence leads to pretty good control capability. However, for attributes with fine-grained levels and relatively ambiguous boundaries, this simple approach leaves room for improvement in control accuracy. To enhance the model's control capabilities, we introduce classifier-free guidance (CFG) \cite{ho2021classifier}, which is initially used in score-based generative models and performs well in aligning conditional input and results. We investigate CFG in the ST generation stage. 

Specifically, during the training of the style LM, we randomly replace the controlling labels with a special empty control token with a probability of $p=0.15$. During inference, for each position $i$, we apply correction to the logit value of style-rich token $s_i$ with the formula
\begin{equation}
    \begin{aligned}
        & \log \hat{\mathrm{P}}(s_t^i|\mathbf{s}_{<t}, \mathbf{s}_t^{<i},\tau,c; \theta_s) \\
        & = \gamma \log \mathrm{P}(s_t^i|\mathbf{s}_{<t}, \mathbf{s}_t^{<i},\tau,c; \theta_s) \\
        & + (1-\gamma) \log \mathrm{P}(s_t^i|\mathbf{s}_{<t}, \mathbf{s}_t^{<i},\tau, \varnothing; \theta_s)
    \end{aligned}
\end{equation}
where $\tau$, $c$, and $\gamma$ represent text (phonemes), control labels, and the guidance scale, respectively. The re-calculated logit is then used for calculating the probability for sampling with the softmax function. Appropriate CFG scales improve the style coherence between the generated speech and the fine-grained control labels, boosting the control capability of the model to some extent. Note that we conduct only CFG on discrete control labels but not on speaker embeddings.

\section{Experiments}
\begin{table*}[ht]
    \centering
    \small
    \caption{Extracting tools and binning strategies for different attributes.}
    \vspace{0.5em}
    \label{tab:label}
    \begin{tabular}{lccccc}
        \toprule
        Attribute & Extracting Tool & Lower Bound & Upper Bound & Bin Number & Interval Boundaries \\
        \midrule
        Gender & w2v2-age-gender\footnotemark[1] & 0.0 & 1.0 & 4 & [0.35, 0.5, 0.65] \\
        Age & w2v2-age-gender & 0 & 100 & 10 & Equidistant \\
        Arousal, Dominance, Valence & w2v2-emotion\footnotemark[2] & 0.2 & 0.8 & 7 & Equidistant \\
        Pitch Mean & DataSpeech &  45.0 & 320.0 & 10 & Equidistant \\
        Pitch Std & DataSpeech & 0.0 & 132.0 & 10 & Equidistant \\
        SNR & DataSpeech & -9.16 & 77.13 & 10 & Equidistant \\
        C50 & DataSpeech & 0.0 & 25.0 & 10 & Equidistant  \\
        \bottomrule
    \end{tabular}
\end{table*}

\begin{table}[ht]
    \small
    \centering
    \caption{Hyper-parameters of different modules of our approach.}
    \vspace{0.5em}
    \label{tab:hyperparameters}
    \begin{tabular}{c|lc}
    \toprule
    Model & \multicolumn{2}{c}{Hyperparameter}   \\ 
    \midrule
    \multirow{5}{*}{Style MAE} & Encoder Layers & 12 \\
    & Decoder Layers & 2 \\
    & Hidden Dimension & 768 \\
    & Mask Probability & 0.75 \\
    & Fbank Channels & 128 \\
    \midrule
    \multirow{6}{*}{\tabincell{c}{Style LM \& \\ Acoustic LM}} 
    &Global Layers         &    20     \\
    &Local Layers & 6 \\
    &Hidden Dim                   &   1,152  \\    
    &Global Attention Heads           &  16   \\  
    &Local Attention Heads           &  8   \\  
    &FFN Dim              &  4,608   \\    
    \bottomrule
    \end{tabular}
\end{table}

\begin{table*}[t]
    \small
    \centering
    \caption{Comparing reconstructed speech from phonemes and ground truth style-rich tokens to original speech, compressed speech and zero-shot TTS results.}
    \vspace{0.5em}
    \label{tab:res1}
    \begin{tabular}{lccc|ccc}
    \toprule
    & \multicolumn{3}{c|}{LibriTTS} &  \multicolumn{3}{c}{Gigaspeech} \\
    \midrule
    Method & SIM & MCD & UTMOS  & SIM & MCD & UTMOS \\ 
    \midrule
    GT. & / & / & 4.06 $\pm$ 0.05 &  / & / & 3.47 $\pm$ 0.10 \\
        GT. + Codec  & 0.94 & 1.98 & 3.43 $\pm$ 0.06 &  0.91 & 2.21 & 2.87 $\pm$ 0.09 \\
    YourTTS & 0.91 & 6.12 & 3.61 $\pm$ 0.09 &  0.85 & 6.72 & 2.33 $\pm$ 0.09 \\
    XTTS-V2  & 0.91 & 5.96 & 3.68 $\pm$ 0.08  & 0.87 & 6.48 & 3.26 $\pm$ 0.10 \\
    Acoustic LM + GT Style & 0.90 & 3.19 & 3.63 $\pm$ 0.05  & 0.86 & 3.68 & 3.24 $\pm$ 0.08 \\
    \bottomrule 
    \end{tabular}
\end{table*}

\subsection{Dataset and style attributes labeling}
For training the style MAE, We adopt a mixed corpora of Gigaspeech-xl \cite{chen2021gigaspeech} and Librispeech \cite{panayotov2015librispeech}. We use Gigaspeech-xl solely for training the style LM, and use high-quality LibriTTS \cite{zen2019libritts} with a relatively small scale for training the acoustic LM. For evaluation, we randomly pick small sets of samples respectively from LibriTTS (184 samples), GigaSpeech (173 samples), and a dialogue dataset, DailyTalk \cite{lee2023dailytalk} (201 samples), to evaluate the models' performance across different data domains. 

To train the style LM, we need to label the different attributes of the data. We utilize multiple annotation tools to extract continuous values or classification probabilities for different speech attributes, and split them into different bins by performing division within an upper and lower boundary that covers most of the data to obtain the discrete control labels. Information about labeling tools and interval splitting strategies are summarized in table~\ref{tab:label}.

\subsection{Metrics}

Our evaluation of model performance primarily consists of speech naturalness, content accuracy, speaker similarity, speech reconstruction quality, and control accuracy. We adopt different objective metrics for evaluation. For speech naturalness, we adopt UTMOS \cite{saeki2022utmos} to predict the MOS score of each sample and report mean values and 95\% confidence intervals for each test set. For content accuracy, we use Whisper large-v3 \cite{radford2022robust} to transcribe the speech and calculate the word error rate (WER) against the ground truth text. For speaker similarity, we compute cosine similarity on speaker embedding extracted by wavlm-base-plus-sv \footnote{\url{https://huggingface.co/microsoft/wavlm-base-plus-sv}}. For reconstruction quality, we calculate MCD between generated and ground truth speech with tools provided in fairseq \footnote{\url{https://github.com/facebookresearch/fairseq/blob/main/examples/speech_synthesis/docs/ljspeech_example.md\#mcdmsd-metric}}. For control accuracy, we use the annotation tools to extract attribute labels and compute percentage accuracy with ground truth labels. Considering the challenges of achieving precise control with fine-grained labels, we make some relaxation that results differing from the ground truth attribute label by one bin are also considered correct for \textit{age}, \textit{SNR} and \textit{C50}, and are considered half correct (taken as 0.5 correct samples) for emotion and pitch labels.

We also conduct subjective evaluations and report mean-opinion-scores of speech naturalness (MOS-Q), style alignment with control labels (MOS-A), and timbre similarity with the reference speaker (MOS-S). We invite 10 individuals with experience in TTS research as participants for our subjective evaluation. For each experiment setting, we select 8 samples for evaluation. The participants rate scores on 1-5 Likert scales, and we report mean scores with 95\% confidence intervals. For MOS-A, considering that the original VAD labels are difficult to understand, we converted the VAD label combinations into emotional intensity levels (such as \textit{flat}, \textit{neutral}, or \textit{highly expressive}) or typical emotional categories (such as \textit{happy}, \textit{angry}, or \textit{sad}) corresponding to those combinations. 

\subsection{Implementation details}

In table~\ref{tab:hyperparameters}, we illustrate the model hyper-parameters of the style MAE and two language models in our approach. For codec, we train an EnCodec \cite{defossez2022high} model for 16k audio, with 8 quantization levels, a codebook size of 1024, and a downsampling rate of 320. We use the first 3 quantization levels only. We also use 3 RVQ codebooks for style-rich tokens.

\subsection{Results and analysis}

\subsubsection{Reconstruct speech style from style-rich tokens and phonemes}
\label{sec:recon}

To validate that our style-rich tokens encapsulate rich voice style information, we reconstruct speech from phonemes and ground truth (GT) style-rich tokens, and compare them with original speech, compressed speech from the codec, and zero-shot TTS results. We use YourTTS \cite{casanova2022yourtts} and XTTS-V2 \cite{casanova2024xtts} as representative zero-shot TTS systems for comparison. The results on LibriTTS and Gigaspeech are shown in table~\ref{tab:res1}. For results on both test sets, our model achieves comparable UTMOS to recent zero-shot TTS systems, and even higher than the compressed GT from the codec. This demonstrates the reliability of our model in terms of speech naturalness. Besides, our model achieves comparable speaker similarity with zero-shot TTS systems, indicating that the style-rich tokens contain rich speaker information for speech synthesis. Moreover, the reconstruction results have significantly lower MCD than zero-shot TTS, proving that it is closer to the original audio in terms of prosody and other style information like acoustic environment, which further validates the effectiveness of our style MAE.

\begin{figure}[t]
    \centering
    \includegraphics[width=0.31\textwidth]{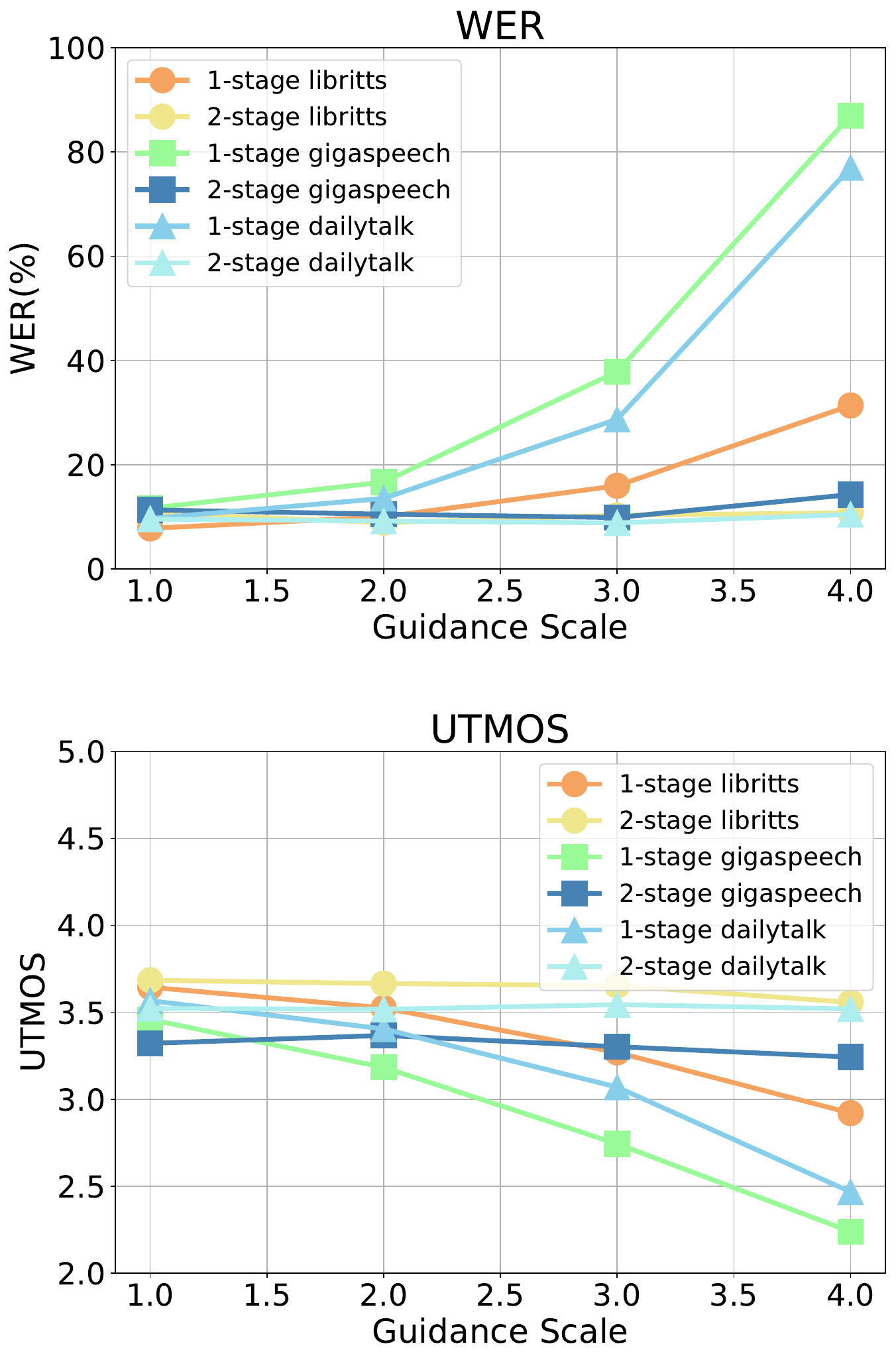}
    \caption{WER and UTMOS on different guidance scales.}
    \label{fig:wer_mos}
    \vspace{-1em}
\end{figure}

\begin{figure*}[tb]
    \centering
    \includegraphics[width=0.79\textwidth]{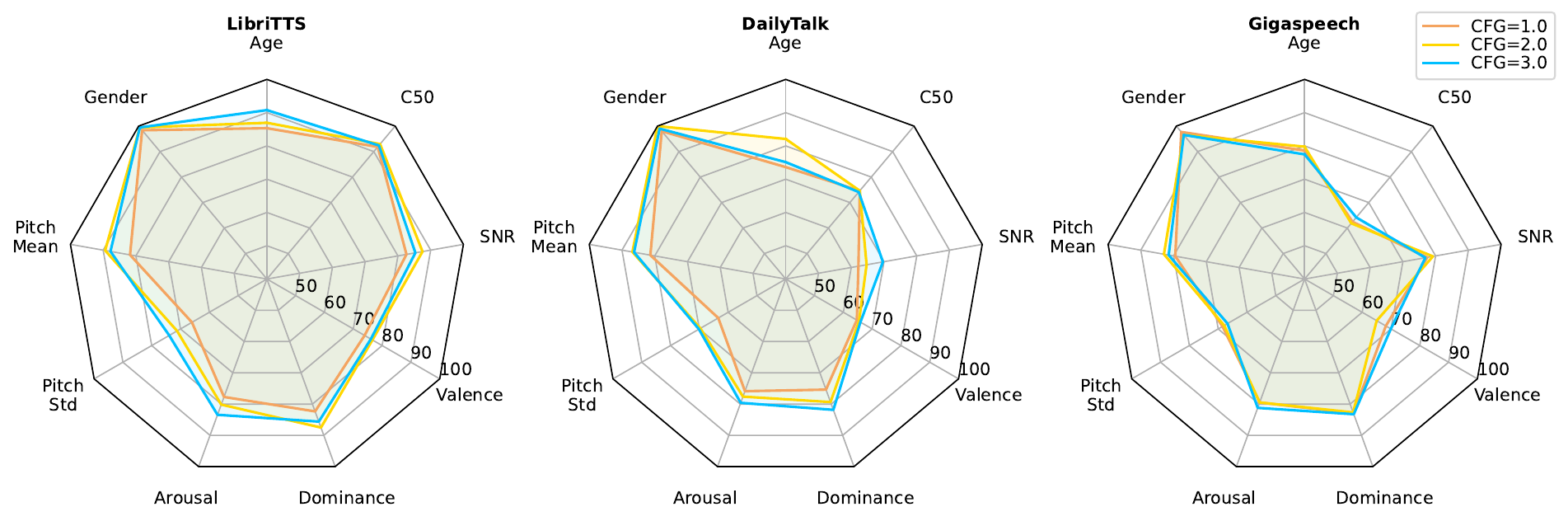}
    \caption{Control accuracy of the two-stage controllable TTS with discrete labels under different CFG scales. The coordinate range is also set to 40-100.}
    \label{fig:cfg}
    \vspace{-1em}
\end{figure*}

\begin{figure*}[t]
    \centering
    \includegraphics[width=0.79\textwidth]{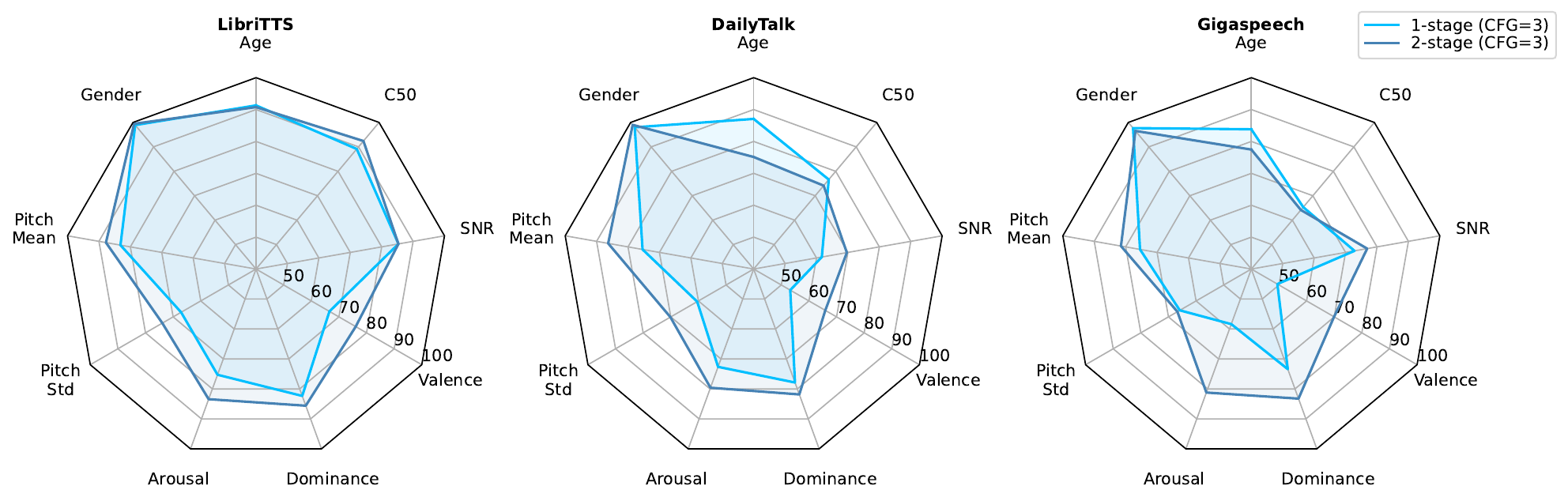}
    \caption{Control accuracy of the one-stage and two-stage controllable TTS with discrete labels under a CFG scale of 3.0. The coordinate range is set to 40-100 for the more apparent differences.}
    \label{fig:stage}
    \vspace{-1em}
\end{figure*}

\footnotetext[1]{\url{https://github.com/audeering/w2v2-age-gender-how-to}}
\footnotetext[2]{\url{https://github.com/audeering/w2v2-how-to}}

\subsubsection{Controllable TTS with discrete labels}
\label{sec:discrete}

In this section, we evaluate the performance of our controllable TTS system with solely discrete labels. Considering the differences in control interfaces, target attributes and training data, it is difficult to directly compare our model with previous controllable TTS systems. To validate the effectiveness of our two-stage design, we train a one-stage model as the baseline, which generates codec tokens from phonemes and control labels directly. We use LibriTTS to train the one-stage model, which is the same as training the acoustic LM. Due to the sheer magnitude of their quantity, traversing all possible attribute combinations is not feasible. Furthermore, the correlation among attributes may render certain combinations of labels impossible or difficult to achieve. Therefore, we use label combinations extracted from ground truth speech for control and evaluation.

We first consider the content accuracy and naturalness of the TTS systems. We illustrate the WER and UTMOS values of the two models under different CFG scales in figure~\ref{fig:wer_mos}. It can be seen that for the one-stage model trained on LibriTTS, as the CFG scale increases, the word error rate rises and UTMOS declines, especially on out-of-domain test sets of Gigaspeech and DailyTalk, manifesting significant degradation in content accuracy and naturalness. This indicates the instability of the one-stage model trained on small, high-quality datasets when subjected to an increased CFG scale, making it difficult to balance control capabilities with speech quality. On the other hand, the two-stage model with the first stage trained on large corpora exhibits good and stable content accuracy and naturalness with growing CFG scales. This proves that the first stage trained on extensive data helps in enhancing the content robustness of controllable TTS.

In figure~\ref{fig:cfg}, we illustrate the control accuracies of the two-stage model under different CFG scales. We can see that the effect of CFG varies for different attributes. For gender attributes with fewer categories and significant differentiation, the presence or absence of CFG shows no clear impact and the model achieves good control performance in both cases. However, for fine-grained attributes like \textit{arousal} and \textit{pitch mean}, appropriate CFG scales can benefit control accuracy, especially on LibriTTS and DailyTalk test sets. This indicates that CFG helps in the precise control of fine-grained attributes. Meanwhile, we find that larger CFG scales are not always beneficial. For some attributes, control accuracy initially increases before subsequently declining as the scale rises. We speculate that this may be due to larger scale values causing distortion in the generated speech, similar to the phenomenon observed with CFG in score-based models.

We further evaluate the control ability of the models. In figure~\ref{fig:stage}, we compare the control accuracies of the one-stage and the two-stage model under a CFG scale of 3.0. It can be seen that the one-stage model has some advantages in \textit{age} control, while the two-stage model achieves comparable or superior control over other attributes. The two-stage model shows significant advantages in emotion control and average pitch, and it also achieves better accuracy over pitch variation and \textit{SNR} on part of the test sets. This indicates that compared to the one-stage model trained on high-quality corpora with limited scale, the two-stage model with the first stage trained with extensive data boosts modeling diverse pitch and acoustic conditions.

\subsubsection{Controlling pitch and emotion with a reference speaker}

\begin{table*}[]
    \centering
    \small
    \caption{Results of controllable TTS combining speaker embedding, pitch and emotion labels.}
    \label{tab:res_3}
    \vspace{0.5em}
    \begin{tabular}{lcccccccc}
    \toprule
    Test set & Model & CFG Scale & WER(\%) & SIM & Aro. & Dom. & Val. & UTMOS  \\
    \midrule
    \multirow{6}{*}{Gigaspeech}  & \multirow{3}{*}{1-stage} & 1.0 & 13.0 & 0.85 & 69.1 & 74.9 & 63.0 & 3.33 $\pm$ 0.08 \\
& & 2.0 & 12.1 & 0.85 & 73.1 & 77.7 & 67.1 & 3.30 $\pm$ 0.08 \\
& & 3.0 & 13.6 & 0.86 & 70.5 & 75.7 & 62.1 & 3.27 $\pm$ 0.07 \\
    \cmidrule{2-9} 
    & \multirow{3}{*}{2-stage} & 1.0 & 12.5 & 0.86 & 76.9 & 76.3 & 68.5 & 3.24 $\pm$ 0.09 \\
& & 2.0 & 14.2 & 0.85 & 78.0 & 78.6 & 68.5 & 3.26 $\pm$ 0.09 \\
& & 3.0 & 13.9 & 0.86 & 76.0 & 80.9 & 65.6 & 3.24 $\pm$ 0.09 \\
    \midrule
    \multirow{6}{*}{DailyTalk}  & \multirow{3}{*}{1-stage} & 1.0 & 14.3 & 0.82 & 65.7 & 71.6 & 58.5 & 3.28 $\pm$ 0.07 \\
& & 2.0 & 13.1 & 0.82 & 66.7 & 71.6 & 59.5 & 3.24 $\pm$ 0.07 \\
& & 3.0 & 14.9 & 0.82 & 68.9 & 75.1 & 59.0 & 3.18 $\pm$ 0.07 \\
    \cmidrule{2-9} 
    & \multirow{3}{*}{2-stage} & 1.0 & 9.5 & 0.80 & 73.9 & 78.6 & 62.7 & 3.50 $\pm$ 0.07 \\
& & 2.0 & 9.5 & 0.80 & 76.1 & 81.6 & 64.4 & 3.53 $\pm$ 0.07 \\
& & 3.0 & 9.3 & 0.80 & 79.6 & 83.3 & 63.7 & 3.51 $\pm$ 0.07 \\
    \bottomrule
    \end{tabular}
\end{table*}

In this section, we present the results that alternate the timbre-related labels including \textit{age} and \textit{gender} with speaker embedding from WeSpeaker \cite{wang2023wespeaker} to achieve control over emotion attributes with a specified reference speaker, and investigate the emotion control capability of the model. The pitch and acoustic condition labels are kept in the condition sequence. We present results on Gigaspeech and DailyTalk in table~\ref{tab:res_3}. It can be seen that our model achieves decent speaker similarity on both test sets as well as comparable control accuracy to the discrete-label-only paradigm over emotion. This indicates the effectiveness of our model in controlling emotion for a specified speaker. Moreover, compared to fully-discrete-label controlling, the one-stage model shows better content robustness with growing CFG scale in this setting, and the one-stage and two-stage models exhibit comparable performance in content accuracy and speaker similarity. Despite this, the two-stage model retains advantages in control over the emotional attributes, demonstrating that the ST generation model trained on an extensive dataset remains advantageous in modeling pitch-related stylistic information in this setting. 

\subsubsection{Subjective evaluation on model performance}

\begin{table}[t]
    \centering
    \small
    \caption{Subjective evaluation results.}
    \label{tab:res_mos}
    \vspace{0.5em}
    \resizebox{0.48\textwidth}{!}{
    \begin{tabular}{lcccc}
    \toprule
    Model & CFG Scale & MOS-Q & MOS-A & MOS-S  \\
    \midrule
    \multicolumn{5}{l}{Control with discrete labels} \\
    \midrule
    \multirow{3}{*}{1-stage} & 1.0 & 4.11 $\pm$ 0.15 & 3.99 $\pm$ 0.18 & / \\ 
    & 2.0 & 3.81 $\pm$ 0.17 & 3.98 $\pm$ 0.15 & / \\
    & 3.0 & 2.89 $\pm$ 0.20 & 3.45 $\pm$ 0.19 & / \\
    \midrule
    \multirow{3}{*}{2-stage} & 1.0 & 4.14 $\pm$ 0.19 & 3.93 $\pm$ 0.18 & / \\
    & 2.0 & 4.01 $\pm$ 0.15 & 4.20 $\pm$ 0.19 & / \\
    & 3.0 & 4.18 $\pm$ 0.18 & 4.20 $\pm$ 0.16 & / \\
    \midrule
    \multicolumn{5}{l}{Control with speaker embeddings and emotion labels} \\
    \midrule
    \multirow{3}{*}{1-stage} & 1.0 & 3.96 $\pm$ 0.16 & 3.84 $\pm$ 0.23 & 3.67 $\pm$ 0.16 \\ 
    & 2.0 & 3.90 $\pm$ 0.15 & 3.90 $\pm$ 0.20 & 3.59 $\pm$ 0.19  \\
    & 3.0 & 3.70 $\pm$ 0.17 & 3.88 $\pm$ 0.17 & 3.40 $\pm$ 0.19 \\
    \midrule
    \multirow{3}{*}{2-stage} & 1.0 & 3.98 $\pm$ 0.17 & 4.06 $\pm$ 0.18 & 3.56 $\pm$ 0.17 \\
    & 2.0 & 4.13 $\pm$ 0.15 & 4.23 $\pm$ 0.16 & 3.68 $\pm$ 0.18 \\
    & 3.0 & 3.91 $\pm$ 0.15 & 4.28 $\pm$ 0.14 & 3.52 $\pm$ 0.18 \\
    \bottomrule
    \end{tabular}
    }
\end{table}

Table~\ref{tab:res_mos} presents the results of our subjective evaluations. As shown, the two-stage model demonstrates comparable or superior MOS-A to the one-stage model, indicating its superior control capabilities. Additionally, an appropriate CFG scale leads to better control performance. Meanwhile, for the one-stage model trained with a small dataset, increasing the CFG scale while using only the labels as the control signal leads to a decrease in MOS-Q. These results align with the conclusions reflected by the objective metrics.

\section{Correlation among control attributes}
\label{appendix:corr}

In fact, the information contained among different attributes may overlap, manifesting as correlations between labels. Certain high-level attributes can be reflected in lower-level acoustic properties. For example, attributes related to speaker timbre, such as age and gender, are closely linked to average pitch, while emotion is closely related to pitch variation. Additionally, the limited performance of the annotation tools may also lead to significant correlations
among different emotional dimensions. For example, the three dimensions of arousal, dominance, and valence are theoretically orthogonal. However, we observe that the distributions of arousal and dominance extracted by the model exhibit a strong linear correlation. 

Due to the correlation among different attributes, using control signals that contain conflict information may lead to sub-optimal speech quality and control capability. We showcase examples on our demo page where conflicting control signals lead to degraded control performance. To achieve better control accuracy and content quality, we can restrict the ranges of low-level attributes with desired high-level attribute labels, thereby avoiding information conflicts. A straightforward solution is a statistical approach, where we can calculate the conditional distributions of \textit{pitch mean} and \textit{pitch std} given other labels on the training dataset, and sample labels from the distribution. Another solution is a learning-based method, where we can train label predictors for estimating low-level attributes from the given high-level labels. We train two 3-layer MLPs with a hidden dimension of 160 to predict \textit{pitch mean} and \textit{pitch std} from \textit{age}, \textit{gender}, \textit{arousal}, \textit{dominance} and \textit{valence}. We find that the accuracy of predicting \textit{pitch mean} and \textit{pitch std} can reach around 40\%, while the soft accuracy—considering a label error of no more than 1 as correct—exceeds 80\%.  Once these models finish training, the output probabilities can be used to sample pitch labels.

\section{Conclusion and limitations}

In this paper, we propose an LM-based fine-grained controllable TTS system. We adopt a two-stage generation pipeline, with an autoregressive transformer as the backbone for each stage. We design a masked autoencoder for extracting features with rich style information from the speech and use the discretized feature as the intermediate output of the TTS pipeline. By selectively combining discrete control labels with speaker embeddings, our model supports both generating new speaker timbre while controlling other attributes, and controlling emotion for a specified speaker. Experiments indicate the effectiveness of our model.

Despite that our approach achieves fine-grained control over multiple style attributes, our method and evaluation protocols still suffer from several limitations: 1) Due to the performance limitations of labeling tools, the attribute annotations of the training data may have a bias against real-world values, causing sub-optimal control capabilities of the model. 2) Evaluation with label combinations from real data may present issues of uneven distribution, particularly for attributes with significant distribution bias, such as SNR and C50. Therefore, the evaluation may not fully accurately reflect the model's control capabilities. 3) Due to the performance limitation of the codec, the model performance on SNR and C50 is not yet very satisfactory. 4) Due to their small proportion in the training data, some marginal labels, such as children's timbre and large SNR, may lead to degraded generated audio and diminished control performance. We may explore solutions to these issues in future works.

\clearpage
\bibliographystyle{IEEEtran}
\bibliography{custom}

\end{document}